\documentclass[twocolumn,superscriptaddress,showpacs,nofootinbib,preprintnumbers,secnumarabic,amssymb, nobibnotes, aps, prd]{revtex4-2}
\usepackage[utf8]{inputenc}
\usepackage{graphicx}
\usepackage{latexsym,amsmath,amssymb,amsthm,lmodern,float,url}
\usepackage{natbib}
\usepackage{color}
\usepackage{microtype}
\usepackage{import}
\usepackage{bbold}
\usepackage[plain]{fancyref}
\usepackage{varioref}
\usepackage{slashed}
\usepackage{multirow}
\usepackage{tikz}
\usepackage{scrextend}
\usepackage{braket}
\usetikzlibrary{shapes}
\usetikzlibrary{positioning}
\usepackage[normalem]{ulem}
\usepackage{caption}
\usepackage{subcaption}
\usepackage{units}
\usepackage{qcircuit}

\newcommand{\nstep}{N_{\text{step}}}

\newcommand{\eq}[1]{Eq.~(\ref{eq:#1})}

\usepackage[colorlinks=true,backref=false, linktocpage=true,
citecolor=blue,urlcolor=blue,linkcolor=blue,pdfpagemode=UseOutlines]{hyperref}
\hypersetup{%
  bookmarksnumbered=true,
  pdftitle = {},
  pdfsubject = {},
  pdfauthor = {},
  pdfkeywords = {}
}

\newcommand{\chisq}{\chi^{2} / \text{d.o.f.}}

\begin{document}
\preprint{USTC-ICTS/PCFT-23-25, FERMILAB-PUB-23-0817-T}
\title{Lattice Holography on a Quantum Computer}
\author{Ying-Ying Li}
\email{yingyingli@ustc.edu.cn}
\affiliation{Peng Huanwu Center for Fundamental Theory, Hefei, Anhui 230026, China}
\affiliation{Interdisciplinary Center for Theoretical Study,
University of Science and Technology of China, Hefei, Anhui 230026, China}

\author{Muhammad Omer Sajid}
\email{omersajid9@outlook.com}
\affiliation{Fermi National Accelerator Laboratory, Batavia,  Illinois, 60510, USA}

\author{Judah Unmuth-Yockey}
\email{jfunmuthyockey@gmail.com}
\affiliation{Fermi National Accelerator Laboratory, Batavia,  Illinois, 60510, USA}

\date{\today}

\begin{abstract}
We explore the potential application of quantum computers to the examination of lattice holography, which extends to the strongly-coupled bulk theory regime.
With adiabatic evolution, we compute the ground state of a spin system on a $(2+1)$-dimensional hyperbolic lattice, and measure the spin-spin correlation function on the boundary. 
Notably, we observe that with achievable resources for coming quantum devices, the correlation function demonstrates an approximate scale-invariant behavior, aligning with the pivotal theoretical predictions of the anti-de Sitter/conformal field theory correspondence. 
\end{abstract}

\maketitle
\section{Introduction}
The anti-de Sitter/conformal field theory (AdS/CFT)~\cite{Maldacena:1997re,KLEBANOV199989,Witten:1998qj} correspondence proposes a duality between a strongly-coupled CFT on the $d$-dimensional boundary of a $(d+1)$-dimensional space with hyperbolic isometries, and a weakly-coupled quantum theory of gravity in the bulk, which suggests the study of quantum field theories via their gravitational duals. Yet, exploring the correspondence for strongly-coupled bulk theories remains a challenge.

An alternative method to study quantum field theories is to discretize space-time into a lattice~\cite{RevModPhys.51.659,RevModPhys.55.775,montvay_münster_1994}.
Although sacrificing continuous space-time symmetries, it enables the application of statistical mechanics' numerical techniques to explore strongly-coupled quantum field theories.  The application of lattice field theory to the AdS/CFT correspondence then potentially allows for the non-perturbative study of strongly-coupled bulk dynamics for
boundary physics. Studies combining AdS/CFT and numerical lattice field theory have revealed that even coarse hyperbolic lattices, where only a subset of the original continuous symmetries remain, exhibit power-law correlations supporting an approximate CFT on the lattice's boundary~\cite{Asaduzzaman:2021bcw,PhysRevE.101.022124,PhysRevD.102.034511,PhysRevD.103.094507,Shima:2005vq,wuIsingModelsHyperbolic1996,wuIsingModelsHyperbolic2000,asaduzzaman2023quantum}. 

However, simulating such strongly-coupled systems---
including those in the AdS/CFT context
---becomes costly for large volumes, where large amounts of entanglement can also be involved.
These limitations ultimately stifle classical simulations of such systems, especially in spacetime dimensions greater than two~\cite{PhysRevLett.91.147902}.  On the other hand, a cutting-edge tool poised to enhance theoretical calculations in physics is the quantum computer, which boasts the capacity to enumerate an exponential number of quantum states using linear resources, enabling efficient handling of intricate, highly-entangled states.

An intriguing query then arises: Can quantum computers be utilized in the near term to investigate the holographic principle? The answer to such a question started to be investigated recently~\cite{Jafferis2022} through a many-body simulation of a Sachdev-Ye-Kitaev system~\cite{PhysRevLett.70.3339}.  This exactly soluble model has allowed for insights into the AdS/CFT correspondence through analytical means, and experimental tests ~\cite{PhysRevLett.119.040501,PhysRevB.96.121119,PhysRevA.103.013323,Luo2019}.
In this work, we  investigate the holographic correspondence between a boundary theory and a strongly-coupled spin system situated on a hyperbolic lattice in $2+1$-dimensions where the aforementioned classical simulations will be ultimately limited.
This involves preparing quantum states across various lattice sizes, and measuring boundary observables to detect indications of scale-invariant behavior. Specifically, we focus on the spin-spin correlation function among boundary spins. 

While computing expectation values with a quantum computer via sampling is straightforward, generating the desired quantum state remains a challenging task. One particularly promising near-term algorithm is the adiabatic state preparation (ASP) method~\cite{farhi2000quantum}. Although ASP ensures the preparation of the desired quantum state for a gapped system, accomplishing this with a manageable circuit depth generally remains a significant hurdle. By employing classical simulators, we will show that the quantum state for the hyperbolic lattice theory can be prepared via ASP with a remarkably constant, and achievable circuit depth as the lattice size increases, fostering the realistic test of holography at large lattices with quantum computers.

Section~\ref{sec:ads-cft} introduces the boundary-to-boundary correlation function, our key observable, and our expectations for its behavior. Section~\ref{sec:state-prep} elaborates on the state preparation method used. Sections~\ref{sec:classical-sims} and~\ref{sec:results} present and discuss outcomes from classical simulations of quantum systems on hyperbolic lattices. Finally, Section~\ref{sec:conclusion} offers concluding remarks and future research directions.
 
\section{AdS/CFT correspondence}
\label{sec:ads-cft}
\begin{figure}
    \centering
    \includegraphics[trim={4cm 0.3cm 4cm 0.3cm},width=8.6cm,clip]{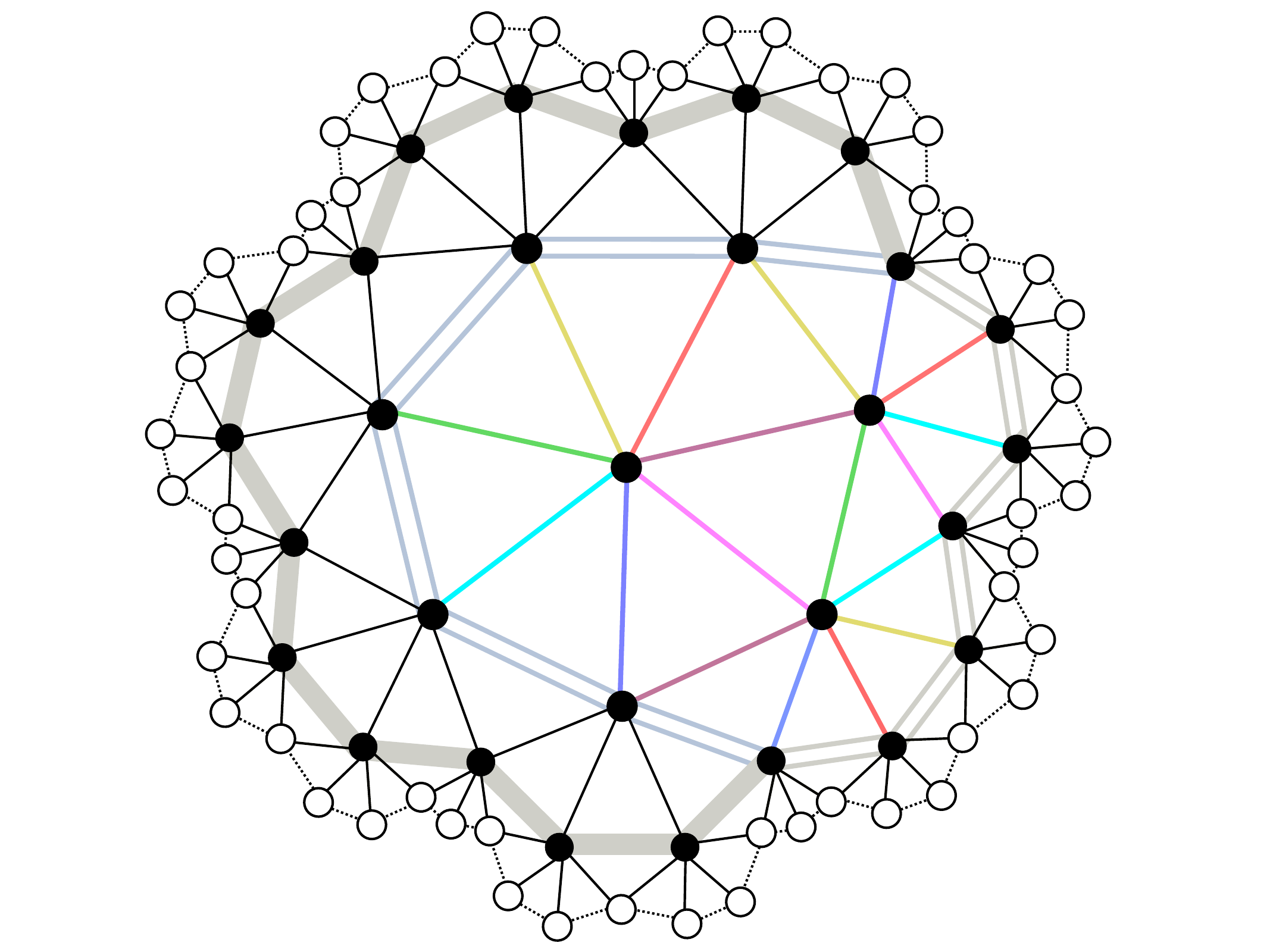}
    \caption{The order-7 triangular lattice with 85 sites.  The lattice boundary is denoted with open circles.  The inner sites outlined by the gray boundary show a smaller volume with 29 sites.  The double-banded edges outline yet a smaller lattice with only 15 sites.  The colored edges demonstrate a pattern for pairs of qubits to be acted on simultaneously by two-qubit gates; that is, the sites bounding all edges of a similar color can be acted on in parallel by two-qubit gates.}
    \label{fig:layer-structure}
\end{figure}
The AdS/CFT correspondence is strictly a relationship between a bulk quantum gravitational theory with AdS shape, and a CFT on the boundary of that spacetime. On the bulk side, we approximate the gravitational theory as a  quantum scalar field theory living on a fixed hyperbolic lattice~\cite{witten1998anti}.  In this case, we work with a discretized and regulated scalar field theory where the lattice spacing is one and bulk fluctuations are given by the transverse Ising model,
\begin{equation}
    \label{eq:ham}
    H = - J_x\sum_{i} X_i - J_{zz} \sum_{\left< i, j\right>} Z_i Z_j,
\end{equation}
with $i$ summing over all lattice sites on the hyperbolic lattice, and $\left< i, j\right>$ being all possible nearest neighbor pairs. $J_x$ and $J_{zz}$ are two couplings quantifying the interaction strength of an external field, and between nearest sites, respectively. We use $X$ and $Z$ to denote the Pauli-$x$ and $z$ operators.  An illustration of the geometry for this lattice can be seen in Fig.~\ref{fig:layer-structure}, for a finite volume. Fig.~\ref{fig:layer-structure} is a projection of the hyperbolic lattice on to the flat plane, 
called the Poincar\'{e} disk.  This causes the edge lengths to vary throughout the lattice when, in fact, all edge lengths are equal in the true hyperbolic lattice.

We consider the bulk theory away from criticality where the system is gapped. This holds true for most pairs of $J_x, J_{zz}$ values. Furthermore we consider the corresponding ground state.  To appreciate the choice of such a generic bulk state,
consider the boundary-to-boundary correlator between boundary site $i$ and site $j$ that are a distance $r$ away on the boundary,
\begin{equation}
\label{eq:2p-corr}
    C(r)=\frac{1}{N_b}\sum^{N_b}_{i}\sum_{j \ni |i-j|=r} \left< Z_i Z_j \right>,
\end{equation}
averaging over all $N_b$ boundary lattice sites. The correlator decays like
\begin{align}
    C(R) \propto e^{-\mu R},
\end{align}
where $R$ is the distance between $i$ and $j$ as measured along the geodesic traversing the bulk, and $\mu$ is the mass gap of the bulk scalar theory for given $J_x$, $J_{zz}$ values~\cite{Hastings2006}. 

Because of the curvature of the lattice, the number of points on the boundary of the lattice grows exponentially with the size of the lattice.  Moreover, for two widely separated boundary points, the geodesic distance traversing the bulk is related to the distance along the boundary by
\begin{align}
    R \approx \alpha L \log(r / L),
\end{align}
where $L$ is the radius of curvature of the lattice, and $\alpha$ is a dimensionless number.  From this relation we can see that, when measured along the boundary, correlations between boundary points go like
\begin{align}
\nonumber
    C(r) \propto e^{-\mu R(r)} &\approx e^{-\mu \alpha L \log(r / L)} \\
    &= \left( \frac{r}{L} \right)^{-\mu \alpha L}.
\end{align}
which is a power-law function of $r$, indicating a conformal theory with no intrinsic scale on the boundary.

The above argument relies on the system possessing a gap in the bulk, and the underlying space being hyperbolic shaped. 
In the next section we discuss how such gapped state preparation is
possible on a quantum computer.

\section{State preparation}
\label{sec:state-prep}
We focus on the $J_{x}$ and $J_{zz}$ values where the bulk theory is gapped. The ground state of \eq{ham} in principle can be prepared via adiabatically evolving the non-interacting ($J_{zz} = 0$) ground state,
$\ket{\psi_{0}} = \ket{+} \otimes \cdots \otimes \ket{+}$, by gradually increasing the strength of $J_{zz}$. 
In practice, the desired ground state cannot be prepared exactly due to the fact that the evolution only proceeds for a finite amount of time, $t$~\cite{adiabatic}.  This finite evolution introduces some error into the calculation, since the adiabatic evolution requires an infinite amount of time to reach the exact ground state.  Moreover, the time evolution itself cannot be carried out faithfully, and must be approximated in some way.  Here we use the Suzuki-Trotter approximation~\cite{trotter,suzuki}.  This method digitizes the continuous-time evolution into a
finite number of steps, $\nstep$, each of which move the system forward by an amount $\Delta t = t / \nstep$.  Therefore, for adiabatic evolution to be successful requires a sufficiently large $\nstep$, and $t$.

The Trottereized time-evolution operator has the form
\begin{align}
\label{eq:trot-ad-ising}
\nonumber
      U = \; &e^{i (\Delta t) J_{x} / 2 \sum_{i} X_{i}} \times \\ \nonumber
    &e^{i n (\Delta t) (\delta J_{zz}) \sum_{\langle i j \rangle} Z_{i} Z_{j}} \times \\
    &e^{i (\Delta t) J_{x} / 2 \sum_{i} X_{i}},
\end{align}
with $n = 1,2,\ldots$, $\nstep$, and $\delta J_{zz} = J_{zz} / \nstep$.
For Eq.~\eqref{eq:trot-ad-ising},   
the depth of a Trotter step---characterized by the number of controlled-NOT (CNOT) gates not run in parallel---is 14, and the total number of CNOT gates is $2\times$(the number of edges). This particular depth is possible because of the seven-fold symmetry of the lattice, which allows for the two-qubit gates within each Trotter step to be organized into seven parallel applications.  These seven layers can be seen in Fig.~\ref{fig:layer-structure} using different colors.

\section{Classical Simulations}
\label{sec:classical-sims}

We examine the validity of ASP by checking the convergences of the average energy $\left< H \right>$ and Eq.~\eqref{eq:2p-corr} in the prepared state using classical simulations of the quantum algorithm.  We perform calculations on four different volumes, $V = 8$, $15$, $29$, and $85$, with the three largest volumes identified in Fig.~\ref{fig:layer-structure}. The simulations are performed on personal laptops, as well as the the OriginQ operating system~\cite{kong2021origin},  with 8000 shots.  We use the \textsc{python} programming language with the \textsc{numpy}~\cite{harris2020array}, \textsc{scipy}, and \textsc{QPanda} packages, \textsc{Mathematica}~\cite{Mathematica}, and IBM's \textsc{qiskit} quantum simulation software~\cite{Qiskit}.  For the largest lattices, we used \textsc{qiskit}'s matrix product state simulator. 

\begin{figure}
    \centering
\includegraphics[width=7.6cm]{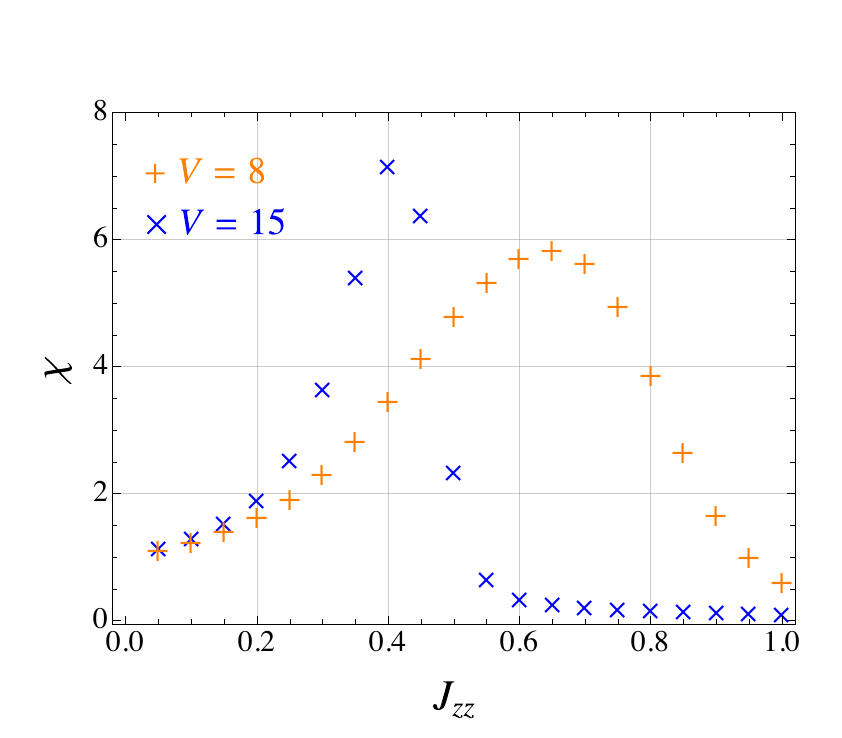}
  \caption{The magnetic susceptibility from exact diagonalization for $V = 8$ and $V = 15$ lattices with a small external field $J_{z} = 10^{-3}$.}
    \label{fig:bulk-suc}
\end{figure}

During ASP we must first ensure that the entire system does not cross a quantum critical point in the adiabatic evolution. As the initial state is chosen to be the disordered state $\ket{\psi_{0}}$, we hope to terminate similarly in the disordered phase which is true for $J_{zz}$ less than a (possible) critical $J_{zz}^{(c)}$.  
To assess the existence of $J_{zz}^{(c)}$, we add a small longitudinal field term to $H$,
\begin{align}
    \delta H = -J_{z} \sum_{i} Z_{i},
\end{align} 
to split a trivial degeneracy in the ordered phase and examine the behavior of the magnetic susceptibility measured over the entire lattice volume $V$, 
\begin{align}
    \chi = \frac{1}{V} ( \langle M^{2} \rangle - \langle M \rangle^{2}), 
\end{align}
with
\begin{align}
    M = \sum_{i} Z_{i}.
\end{align} 
With the small $J_z$ term, a potential phase transition is hinted at by a peak in $\chi$ at the critical coupling $J_{zz}^{(c)}$, as shown in Figure~\ref{fig:bulk-suc} for $V = 8$ and $15$. 
As the volume increases we see a sharpening and growing of the peak, indicating a potential transition when extrapolating to the infinite volume limit.  In the larger volume the peak appears around $J^{(c)}_{zz} \approx 0.4$ which could make adiabatic evolution to larger $J_{zz}$ values difficult.  Since we are interested in a gapped phase regardless, we restrict to $J_{zz} < 0.4$ and disregard the $\delta H$ term going forward.

\begin{figure}
    \centering
  \includegraphics[width=7.6cm]{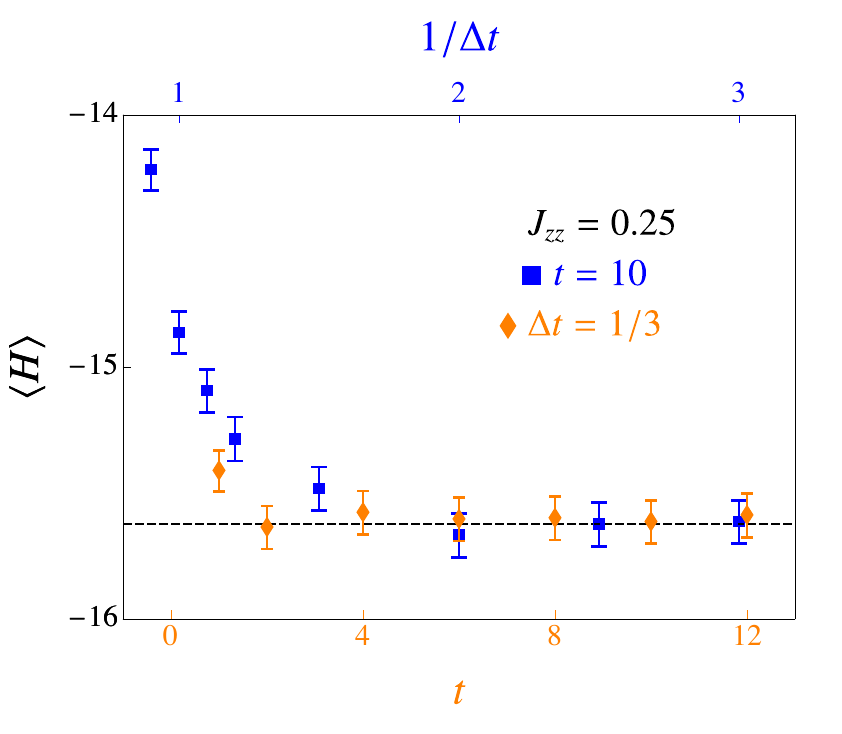}\\
    \caption{Energy convergence from adiabatic evolution using different numbers of Trotter steps for $t=10$ or different $t$ but fixing $\Delta t = 1/3$. The black dashed line is the ground state energy from exact diagonalization for $V=15$.}
    \label{fig:adiabat-nrg-converge}
\end{figure}
\begin{figure}
    \centering
  \includegraphics[width=7.6cm]{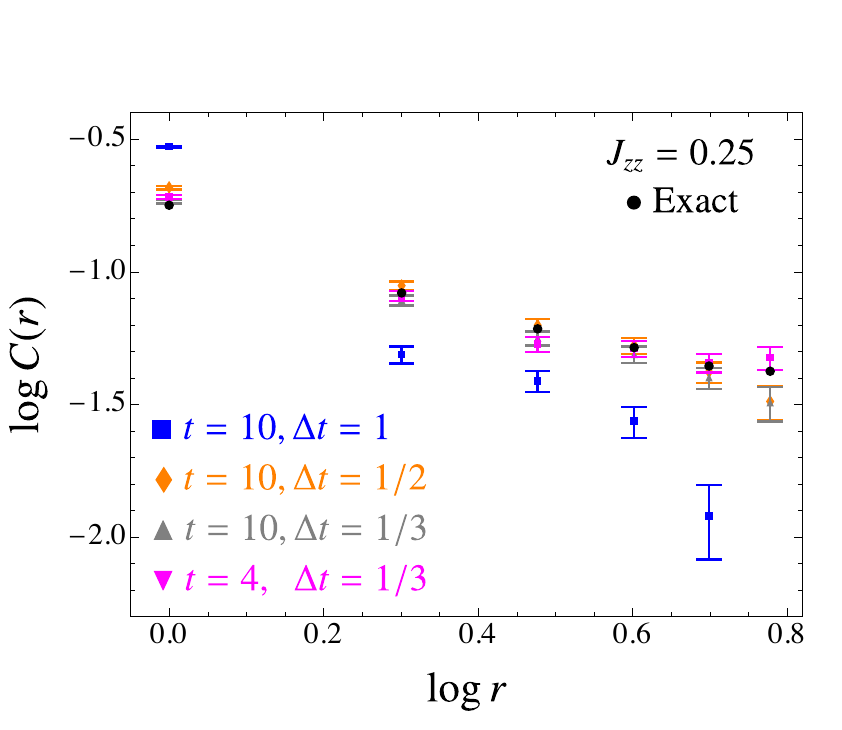}\\
    \caption{Correlation functions on the 15 qubit lattice as a function of boundary distance for different $t$ and $\Delta t$.}
    \label{fig:corr15-convergence}
\end{figure}
With a valid range of $J_{zz}$ values in hand, we can now work to identify ideal values of $t$ and $\nstep$.  To do this we consider a volume of 15 qubits with 60 CNOT gates and $R_{z}$ gates per Trotter step, and we vary $t \in [0, 12]$, and $\nstep \in [10, 200]$. As time increases at fixed $\Delta t$, the algorithm's average energy converges to the exact ground state value for that particular coupling~\cite{PhysRevLett.131.060602}. Figure~\ref{fig:adiabat-nrg-converge} illustrates this convergence, exhibiting agreement in the large-time limit.
Figure~\ref{fig:adiabat-nrg-converge} also depicts the energy convergence in $\Delta t$ at fixed $t$. For sufficiently large $t$ and small $\Delta t$, the average energy approaches the exact value. Additionally, Fig.~\ref{fig:corr15-convergence} shows the convergence of Eq.~\eqref{eq:2p-corr} under similar circumstances.

The practical cost for ASP depends on $\nstep$. From Fig.~\ref{fig:adiabat-nrg-converge}, $\nstep \approx 30$ yields an accuracy consistent with the exact ground state energy, implying a total CNOT depth of $420$---assuming all-to-all connectivity.  This type of connectivity is possible on trapped-ion machines.  For machines with fixed topology such as IBM's \texttt{brisbane} machine, we have compiled a single Trotter step for the 29 qubit case
and found a circuit depth of $\approx 250$, which is realistic for coming quantum devices. We now present results in the next section computed using $t = 10$, and $\Delta t = 1/3$. 

\section{Results}
\label{sec:results}

With the methods under control, we investigate the expected signals mentioned in Sec.~\ref{sec:ads-cft} pertaining to $C(r)$ by performing adiabatic evolution using a digital simulator with classical resources. Using the resultant state prepared for various $J_{zz}$ values, we calculate Eq.~\eqref{eq:2p-corr} between spins on the boundary.  Other thermodynamic observables \cite{Carena:2022hpz} which help articulate the boundary physics can also be calculated and are left for future studies. 
\begin{figure}
    \centering
    \includegraphics[width=7.6cm]{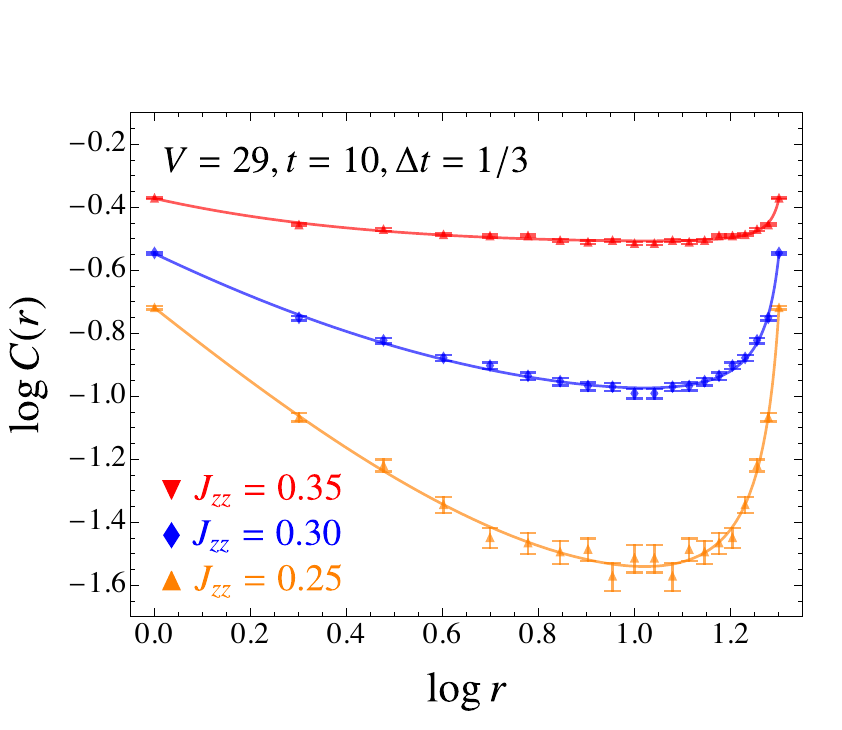}
    \caption{The boundary correlation function on the 29-qubit lattice for three values of $J_{zz}$, along with the corresponding best fit lines using Eq.~\eqref{eq:power}.} 
    \label{fig:29-corr-fit}
\end{figure}
\begin{figure}
    \centering
    \includegraphics[width=7.6cm]{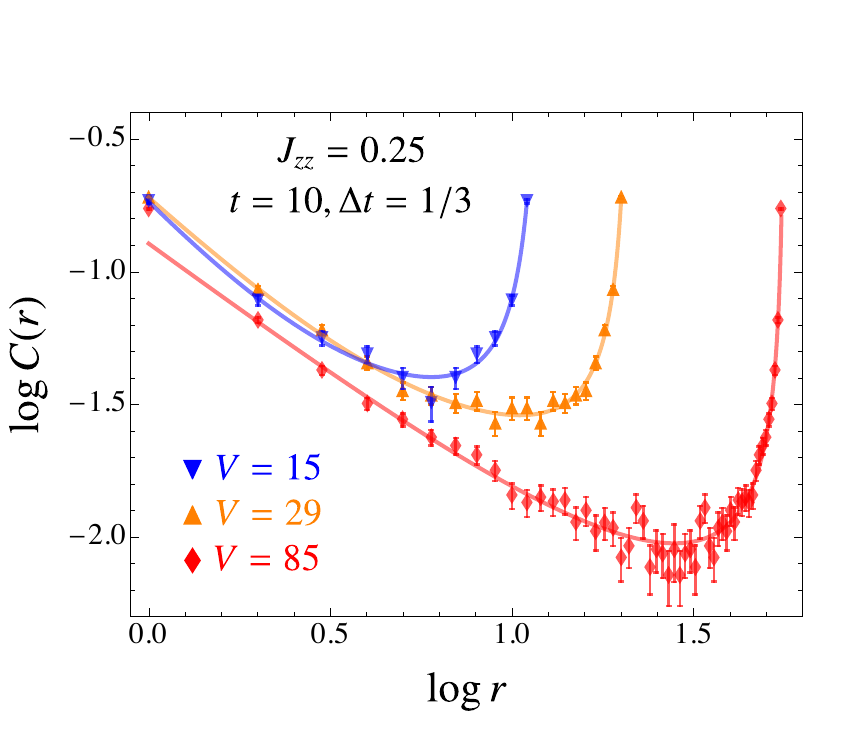}
    \caption{The boundary correlation function for the three volumes considered at a fixed value of $J_{zz} = 0.25$, along with the best fit lines using Eq.~\eqref{eq:power}.} 
    \label{fig:85-corr-fit}
\end{figure}

We fit Eq.~\eqref{eq:2p-corr} to the form
\begin{align}
    C(r) = a \left[ \frac{e^{-b r}}{r^{c}} +  \frac{e^{-b (N_{b}-r)}}{(N_{b}-r)^{c}} \right] + d
\end{align}
where $a,b,c,d$ are free parameters, and the two terms in brackets account for the periodicity of the lattice.  
This form quantifies the exponential decay in the correlation function due to the finite lattice spacing, along with the expected continuum power-law behavior. 
In fact, in all cases, the fitted value for $b$ is consistent with zero, allowing for the exponential dependence to be ignored and instead a pure power-law fit can be used,
\begin{align}
\label{eq:power}
    C(r) = a \left[ \frac{1}{r^{c}} +  \frac{1}{(N_{b}-r)^{c}} \right] + d.
\end{align}
This provides more evidence that the boundary theory is indeed approximately massless.

Figure~\ref{fig:29-corr-fit} demonstrates the results of the fits for three values of the $J_{zz}$ coupling in the disordered phase for $V = 29$.  There, the data is plotted with errorbars, alongside the best fit line.  We find that deep in the disordered phase the fit quality is best, with it degrading as we approach the potential $J_{zz}^{(c)}$ where the bulk theory is strongly-coupled. The $\chisq = 0.5$, $0.8$, and  $1.7$ for $J_{zz} = 0.25$, $0.3$ and $0.35$, respectively, with 17 degrees of freedom.

Figure~\ref{fig:85-corr-fit} similarly shows $C(r)$, however for  three different volumes considered at a fixed $J_{zz} = 0.25$.  Here we again see excellent agreement between data and the fit ansatz, with $\chisq = 5.8/8$, $8.2/17$, and $36.9/50$ for $V = 15$, $29$ and $85$, respectively.  This figure demonstrates the emerging long-distance  power-law behavior visible on larger lattices, exemplified by the linear behavior for the $V=85$ data starting with $r=2$ and ending at $r \approx 21$.

\section{conclusions}
\label{sec:conclusion}
We have investigated the near-term potential to observe scale-invariant behavior of a quantum field theory on the boundary of hyperbolic space using a quantum computer.  This behavior takes the form of near-critical correlation across a range of coupling-constant values. The quality of our results indicate that with modest circuit depths, this behavior could be seen in coming quantum simulations. This quantum capability opens the window to facilitate future study of interesting features of holography such as entanglement entropy, boundary thermodynamics and critical exponents associated with a potential phase transition of the boundary theory. 

Of course, quantum computation in the present day is error-ridden, and future work would also include a repeated analysis diagnosing errors, and implementing error mitigation techniques during the preparation of the scale-invariant state. On the other hand, promising results in engineering error-corrected machines are emerging~\cite{Bluvstein2023}. 
Finally, more complicated models could be considered, such as $\mathbb{Z}_{2}$ gauge theory on similar lattices, or considering spin models on 3+1-dimensional lattices. 

\begin{acknowledgments}
We would like to thank Simon Catterall, and Jay Hubisz, Henry Lamm, Jing Wang, Lei Yu, Chang-Ling Zou for stimulating discussions about this work.

This manuscript has been authored by Fermi Research Alliance, LLC under Contract No. DE-AC02-07CH11359 with the U.S. Department of Energy, Office of Science, Office of High Energy Physics.

This work is supported by the Department of Energy through the Fermilab Theory QuantiSED program in the area of “Intersections of QIS and Theoretical Particle Physics.

We thank the Institute for Nuclear Theory at the University of Washington for its kind hospitality and stimulating research environment. This research was supported in part by the INT's U.S. Department of Energy grant No. DE-FG02- 00ER41132.

Y.-Y. Li and Judah Unmuth-Yockey would like to thank the Munich Institute for Astro-, Particle and BioPhysics (MIAPbP) for their support, which is funded by the Deutsche Forschungsgemeinschaft (DFG, German Research Foundation) under Germany´s Excellence Strategy – EXC-2094 – 390783311.

Y.-Y. L is supported by the NSF of China through Grant
No. 12247103.
\end{acknowledgments}

%
\end{document}